\documentclass[]{aspm}

\articleinfo{}{}{}

\usepackage{verbatim}
\usepackage{amssymb}
\usepackage{amsbsy}
\usepackage{amscd}
\usepackage{amsmath}
\usepackage{amsthm}
\usepackage[mathscr]{eucal}
\usepackage{graphicx,subfigure}% Include figure files

\numberwithin{defn}{section}

\title[Singularities and self-similarity]{Singularities and 
self-similarity in gravitational collapse}

\author[T. Harada]{Tomohiro Harada}

\address{Department of Physics\\ 
Rikkyo University\\
Tokyo\\
Japan}

\email{harada@rikkyo.ac.jp}

\rcvdate{}
\rvsdate{}

\subjclass[2000]{}


\begin{document}

\begin{abstract}
Einstein's field equations in general relativity admit 
a variety of solutions with spacetime singularities.
Numerical relativity has recently revealed the properties of 
somewhat generic spacetime singularities. It has been found that
in a variety of systems self-similar solutions can describe asymptotic
or intermediate behaviour of more general solutions.
The typical example is the convergence to an attractor  
self-similar solution in gravitational collapse.
This is closely related to the cosmic censorship violation 
in the spherically symmetric collapse of a perfect fluid. 
The self-similar solution also plays an important 
role in critical phenomena in gravitational collapse.
The critical phenomena are understood as the intermediate 
behaviour around a critical self-similar solution.
We see that the convergence and critical phenomena
are understood in a unified manner in terms of attractors
of codimension zero and one, respectively, 
in renormalisation group flow.

\end{abstract}

\maketitle

\section{The framework of general relativity}
The essential assumption of general relativity is that the spacetime is given 
by a curved manifold with a metric $ds^{2}=g_{ab}dx^{a}dx^{b}$ of the
Lorentzian signature. $g^{ab}$ denotes the inverse of $g_{ab}$.
The curvature of the spacetime is given by the Riemann tensor $R^{a}_{~~bcd}$.
The metric lifts and lowers the tensor indices.  A vector 
is timelike, spacelike and null if $v^{a}v_{a}<0$, $v^{a}v_{a}>0$ and 
$v^{a}v_{a}=0$, respectively. A hypersurface is called timelike, 
spacelike and null, if its normal vector is spacelike, timelike 
and null, respectively. We use the abstract index 
notation~\cite{wald1983} in this article.

The field equation for the metric is given by Einstein's equations
\begin{equation}
 R_{ab}-\frac{1}{2}g_{ab}R=8\pi T_{ab},
\end{equation}
where $R_{ab}\equiv R^{c}_{~~acb}$ is the Ricci tensor, $R\equiv
R^{a}_{~~a}$ is the scalar curvature and $T_{ab}$ is the stress-energy tensor 
of matter fields.
We adopt the units in which $G=c=1$.
Einstein's equations were proposed so that it has the limit 
to Newtonian gravity in a weak-field and slow-motion regime.
The conservation law
\begin{equation}
\label{conservation_law}
 \nabla _{b}T_{~~a}^{b}=0
\end{equation}
follows from the Bianchi identity, where $\nabla _{b}$ denotes 
the covariant derivative associated with $g_{ab}$. 
See~\cite{wald1983,he1973} for more complete description 
about the formulation of general relativity.

Because of the Lorentzian signature of the metric, 
the spacetime can admit a time function $t$ and 
then ``3+1'' decomposition, i.e. foliation with 
spacelike hypersurfaces labelled by $t$.
This is called time foliation or time slicing of the spacetime
and this enables us to regard Einstein's equations as
the combination of the evolution and constraint equations 
for the induced metric and the extrinsic curvature on the spacelike 
hypersurface.
If the spacetime admits a timelike Killing vector, we can choose
$t$ so that the induced metric does not depend on $t$.
Such a spacetime is called stationary.

General relativity is a self-consistent theory of gravity but not
written in a completely closed form. We need to specify the physics of
matter fields by giving the action or the 
stress-energy tensors of matter fields, which provide 
the source term on the right-hand side 
of Einstein's equations. Given the action of matter fields 
$S_{\rm m}$, the stress-energy tensor is defined as a functional
derivative in the following: 
\begin{equation}
T_{ab}\equiv \frac{2}{\sqrt{-g}}\frac{\delta S_{\rm m}}{\delta g_{ab}},
\end{equation}
where $g$ denotes the determinant of $g_{ab}$.

The simplest source term is vacuum, i.e. $T_{ab}=0$. 
If $T_{ab}=\lambda g_{ab}$ for a constant $\Lambda$,
this is called a cosmological constant.
A perfect fluid is often used, which is given by 
\begin{equation}
T_{ab}=\rho u_{a}u_{b}+p (u_{a}u_{b}+g_{ab}),
\end{equation}
where $u^{a}$ is the four-velocity of the fluid element, satisfying
$u^{a}u_{a}=-1$. This provides a very good approximation for gases, liquids 
and solids in many circumstances. The equation which gives $p$ is
called the equation of state. If $p=0$, 
the perfect fluid is called a dust. 
If $p=\rho/3$, it is called a radiation fluid. 
Another common example is a scalar field, for which the stress-energy 
tensor is given by
\begin{equation}
T^{ab}=\nabla^{a}\phi \nabla^{b}\phi-\frac{1}{2}g^{ab}\nabla_{c}\phi
\nabla^{c}\phi+V(\phi)g^{ab},
\end{equation}
where $V(\phi)$ is called a potential. If $V(\phi)=0$, the 
scalar field is called massless.

The matter fields evolve according to their equations of motion. 
If there is only a single matter component, 
the conservation law~(\ref{conservation_law}) gives the equation of motion.
The metric is a solution of Einstein's equations with the matter fields as
the source. Because of the conservation law, the distribution of matter
fields in space and time cannot be put arbitrarily 
by some ``external force'' which does not contribute to 
the stress-energy tensor.   
In other words, we cannot determine the metric simply after we 
assume the distribution of matter fields as we often do to determine
electric and magnetic fields in electromagnetism. 
We need to solve the metric and matter fields in a consistent manner.
Moreover, Einstein's equations are highly nonlinear with respect to 
$g_{ab}$. These properties make it very difficult 
to get exact solutions in dynamical and generic situations
with or without matter fields.

To obtain the general properties of spacetimes, it is important to know the
general properties of matter fields. Among such conditions for matter
fields are energy conditions, which impose the energy density being positive
in some sense. The strong energy condition is one of them, assuming 
\begin{equation}
R_{ab}\xi^{a}\xi^{b}=8\pi(T_{ab}\xi^{a}\xi^{b} +T^{a}_{~~a}/2)\ge 0,
\end{equation}
for any timelike vector $\xi^{a}$.
This implies $\rho+p\ge 0$ and $\rho+3p\ge 0$ for a perfect fluid.

Almost all known exact solutions to Einstein's equations have been 
obtained under strong assumption on symmetry.
A very powerful method to obtain more or less general solutions 
is to establish numerical solutions. 
This is called numerical relativity in a broad sense. It has 
been achieving great success in recent days, not only in numerical
simulations of relativistic astrophysical phenomena
but also in the numerical experiments of phenomena 
with nonlinearly strong gravity.

\section{Singularities in general relativity}
\subsection{Examples of spacetime singularities}
\begin{figure}[htbp]
\begin{center}
\includegraphics[width=7cm]{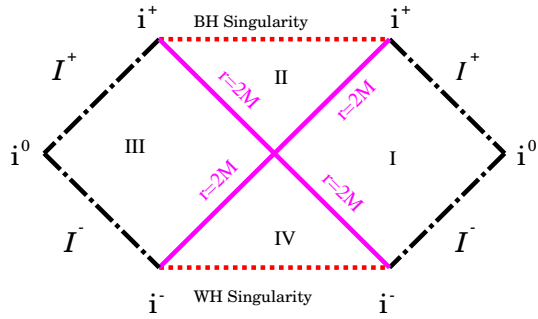}
\caption{\label{schwarzschild_kruskal} The Penrose diagram of 
the Schwarzschild solution for $M>0$.}
\end{center}
\end{figure}

There are exact solutions to Einstein's equations which 
have spacetime singularities. See~\cite{wald1983,he1973,clarke1993} 
for the precise notions of causal structure and spacetime singularities. 
Here we look at several examples.
The first one is the 
Schwarzschild solution, in which the line element 
is written in the following form:
\begin{equation}
 ds^{2}=-\left(1-\frac{2M}{r}\right)dt^{2}+\left(1-\frac{2M}{r}\right)^{-1}dr^{2}+r^{2}(d\theta^{2}+\sin^{2}\theta d\phi^{2}).
\end{equation}
This is the unique solution for spherically symmetric 
vacuum spacetimes. This describes a black hole for $M>0$.
In this case, $r=2M$ is a coordinate singularity, 
corresponding to an event horizon,  
while $r=0$ is a genuine spacetime singularity, towards which
the scalar curvature polynomials tend to diverge.
These features are well understood in the Penrose diagram 
or conformal diagram shown in Fig.~\ref{schwarzschild_kruskal}.
In this figure, the straight line with forty five degrees denotes 
a null ray and the the physical spacetime is compactified through 
the conformal transformation. 
Suppose that an observer is in region I and she tries to send a signal
to future null infinity ${\mathcal I}^{+}$ by emitting a light.
We can see that the 
future-directed outgoing null rays
earlier than $r=2M$, i.e. in region I, can reach ${\mathcal I}^{+}$,
while those later than $r=2M$, i.e. in region II, cannot. 
$r=0$ is described by the future and past boundaries of the spacetime,
which are black hole and white hole singularities, respectively.
No future-directed null ray can emanate from the black hole singularity.

\begin{figure}[htbp]
\begin{center}
\begin{tabular}{cc}
\includegraphics[width=5cm]{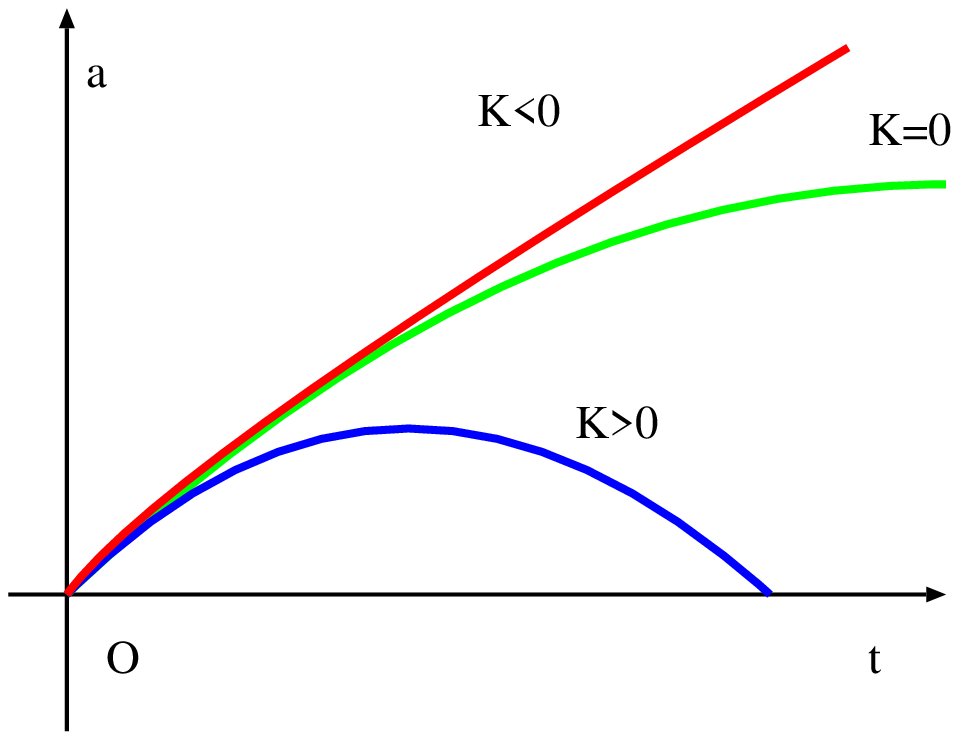}&
\includegraphics[width=5cm]{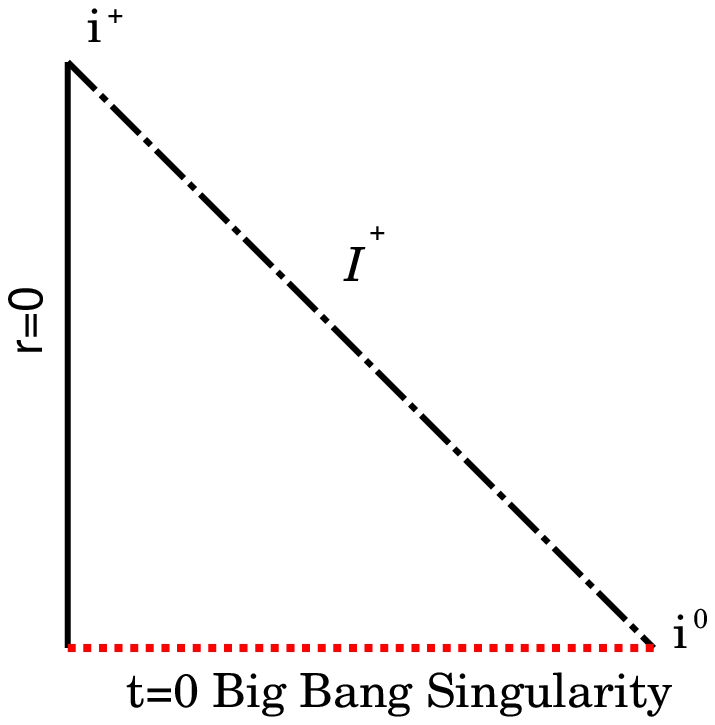}
\end{tabular}
\caption{\label{friedmann} The left and right panels show the evolution
 of the scale factor for different spatial curvatures and 
the Penrose diagram of the flat ($K=0$) Friedmann solution, respectively.}
\end{center}
\end{figure}

The second example is 
the Friedmann solution which describes a 
homogeneous and isotropic universe. 
The line element is then given by 
\begin{equation}
ds^{2}=-dt^{2}+a^{2}(t)\left[\frac{1}{1-Kr^{2}}dr^{2}
+r^{2}(d\theta^{2}+\sin^{2}\theta d\phi^{2})\right], 
\end{equation}
where the scale factor $a=a(t)$ obeys the Friedmann equation
\begin{equation}
\left(\frac{\dot{a}}{a}\right)^{2}=\frac{8\pi}{3}\rho-\frac{K}{a^{2}},
\end{equation}
the matter energy density $\rho$ obeys the energy conservation law
\begin{equation}
\dot{\rho}=-3(\rho+p)\frac{\dot{a}}{a},
\end{equation}
and the dot denotes the derivative with respect to $t$.
$K$ denotes the curvature of the  $t=$const spacelike hypersurface.
Note that the source term for the Friedmann solution must be a perfect fluid.
If the strong energy condition is satisfied, or 
$\rho+p\ge 0$ and $\rho+3p\ge 0$ in this case, 
$a$ then begins with 0 at $t=0$, implying the divergence of 
scalar curvature polynomial where $\rho\to \infty$. 
This is a spacelike singularity and usually called
big bang singularity or initial singularity. Figure~\ref{friedmann} shows 
the evolution of the scale factor and the causal structure.

\begin{figure}[htbp]
\begin{center}
\includegraphics[width=5cm]{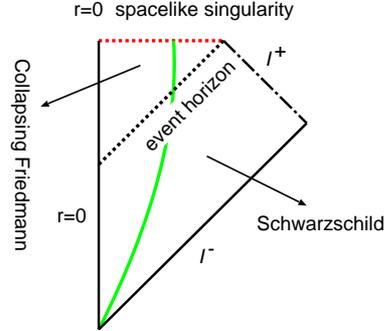}
\caption{\label{os} The Penrose diagram of the Oppenheimer-Snyder solution.}
\end{center}
\end{figure}

The simplest model for gravitational collapse is the Oppenheimer-Snyder 
solution, which describes the complete collapse of a uniform dust ball.
This is given by matching the interior and exterior solutions. 
The interior solution is the time-reversed Friedmann universe with a dust.
The exterior is spherically symmetric and vacuum, which is given by 
the Schwarzschild solution. They are matched on a timelike hypersurface,
which is generated by timelike radial geodesics. 
The Penrose diagram of the resultant spacetime is shown in Fig.~\ref{os}, which is given by cutting and pasting those of 
the Friedmann solution and the Schwarzschild solution. 
We can see that the singularity is spacelike and hidden behind the 
event horizon from the external observer. There exists an achronal, i.e. spacelike or null, three-dimensional
hypersurface such that the whole spacetime is the domain of dependence of this surface.
This surface is called the Cauchy surface and the spacetime with the
Cauchy surface is called globally hyperbolic.

\subsection{Physical significance of singularities}
The geodesic deviation equation implies 
\begin{equation}
\frac{d\theta}{d\tau}=-\frac{1}{3}\theta^{2}-\sigma_{ab}\sigma^{ab}
+\omega_{ab}\omega^{ab}-R_{cd}\xi^{c}\xi^{d},
\end{equation}
along a geodesic congruence, where $\xi^{a}$ is the normalised tangent vector, $\tau$ is an affine 
parameter, and $\theta$, $\sigma_{ab}$
and $\omega_{ab}$ are respectively the expansion, shear and twist. This equation is called 
the Raychaudhuri equation. This implies that the timelike geodesic congruence tends to focus, in other words, $\theta$ tends to diverge to $-\infty$
if $R_{cd}\xi^{c}\xi^{d}\ge 0$ or matter fields satisfy 
the strong energy condition. 

Based on the above properties, it was proved that there is 
at least one incomplete timelike or null geodesic in generic 
expanding universe and generic gravitational 
collapse~\cite{he1973}. This is called singularity theorems.
The geodesic incompleteness implies 
the existence of spacetime singularities.
This is a great achievement in the studies of classical general relativity.
However, singularity theorems would not reveal 
the properties of generic singularities.
Then, 
we see two important conjectures on this issue.  

We first consider initial singularities. 
The Kasner solution is given by
\begin{equation}
ds^{2}=-dt^{2}+ t^{p_{1}}dx^{2}+t^{p_{2}}dy^{2}+t^{p_{3}}dz^{2},
\end{equation}
where $p_{1}$, $p_{2}$ and $p_{3}$ are three indices,
satisfying $p_{1}+p_{2}+p_{3}=p_{1}^{2}+p_{2}^{2}+p_{3}^{2}=1$ and $p_{1}<0<p_{2}<p_{3}$. This solution
describes a homogeneous and anisotropic vacuum universe and 
has initial singularity at $t=0$. The dynamics of 
the Bianchi type IX homogeneous universe was shown to be
the successive series of Kasner regimes replacing the indices one 
another~\cite{blk1970}.
In each regime the solution is well approximated by the Kasner solution but 
the different regime has different indices $(p_{1},p_{2},p_{3})$. 
The transition is oscillatory and chaotic. Subsequently,
it was conjectured that 
the process of approach to initial singularity in the general
inhomogeneous case tends to be local, oscillatory and chaotic and 
the dynamics of nearby observers decouple 
from each other near the singularity~\cite{bkl1982}.
This conjecture is called the Belinski-Khalatnikov-Lifshitz conjecture and 
has been recently strongly supported by numerical relativity
experiment of initial singularity with no symmetry~\cite{garfinkle2004}.

As for spacetime singularities formed in gravitational collapse, 
the cosmic censorship conjecture was proposed by Penrose. 
For the weak version, he conjectured,
``A system which evolves, according to classical general relativity
 with reasonable equations of state, from generic non-singular initial
 data on a suitable Cauchy-hypersurface, does 
not develop any spacetime singularity which is visible from infinity''~\cite{penrose1969}.
For the strong version, he conjectured,
``$\cdots$ a physically reasonable classical spacetime $M$ ought to have
the property $\cdots$ $M$ is globally hyperbolic $\cdots$''~\cite{penrose1979}.
A singularity which is censored by this conjecture is called a naked singularity. 
In other words, the cosmic censorship claims that there 
is no naked singularity in 
physical spacetimes.

Obviously, there are undefined terms in this conjecture, such as 
reasonable equations of state and generic initial data.
The cosmic censorship is a basic assumption to prove theorems
on the properties of black holes, such as no bifurcation, area increase
and the existence of an event horizon outside or coinciding to 
an apparent horizon~\cite{he1973}.
The proof for the cosmic censorship is still very limited. On the other hand, 
it has been revealed that there are a lot of solutions of Einstein's
equations, which satisfy 
energy conditions and have naked singularities.
We do not know how to apply known physics 
at spacetime singularities. Hence, if there is a naked 
singularity, it would spoil the 
the predictability for our future within classical theory.
On the other hand, naked singularities may be regarded as a window into 
physics beyond general relativity~\cite{hn2004}.
See~\cite{joshi2000,hin2002,harada2004} for naked-singular solutions 
in gravitational collapse and 
possible physical processes in naked singularity formation.

\section{Gravitational collapse and self-similar solutions}
\subsection{Self-similar solutions}
General relativity has no characteristic scale of its own, which 
implies the existence of self-similar 
solutions. Self-similar solutions are easier 
to obtain than more general solutions
because partial differential equations reduce to 
ordinary differential equations for spherically symmetric self-similar spacetimes.
For self-similar solutions, the energy density $\rho$, for example, is written as
$\rho(t,r)=t^{-2}f(r/t)$ with appropriate time and radial coordinates
$t$ and $r$. 
It is found that in some spatially homogeneous models 
self-similar solutions can describe the asymptotic behaviour of 
more general solutions~\cite{we1997}. 
It was also conjectured that spherically symmetric fluctuations
might naturally evolve via Einstein's equations from a complex 
initial conditions to a self-similar form~\cite{carr1993}.
See~\cite{cc2007} for a recent review of self-similar 
solutions in general relativity.

More precisely, if a vector field $\xi$ satisfies
\begin{equation}
{\mathcal{L}}_{\bf\xi} g_{ab} =2 g_{ab},
\end{equation}
this is said to be a homothetic Killing vector.
If there exists a homothetic Killing vector in a spacetime,  
this spacetime is said to be self-similar or homothetic.
For a spherically symmetric self-similar spacetime,
introducing coordinates $(t,r)$ such that 
\begin{equation}
 \xi=t\frac{\partial }{\partial t}+r\frac{\partial}{\partial r},
\end{equation}
a nondimensional quantity $Q$ satisfies
\begin{equation}
Q(t,r)=Q(a t, a r), 
\end{equation}
for any $a>0$ and hence $Q=Q(r/t)$. The line element is then 
given by 
\begin{equation}
ds^{2}=-e^{\sigma(z)}dt^{2}+e^{\omega(z)}dr^{2}
+r^{2}S^{2}(z)(d\theta^{2}+\sin^{2}\theta d\phi^{2}), 
\end{equation}
where $z\equiv \ln|r/(-t)|$.
On the other hand, if there exists a positive $\Delta$ such that 
$Q(t,r)=Q(e^{n\Delta} t, e^{n\Delta} r)$ holds only for 
$n=0,\pm 1, \pm 2, \cdots$, this is called discretely self-similar. 

To have self-similar spacetimes, matter fields are strongly
restricted. Both a perfect fluid with the equation of state 
$p=k\rho$ and a massless scalar field $\phi$ are still compatible with 
self-similar spacetimes.
For self-similar spacetimes, Einstein's equations reduce to 
a set of ordinary differential equations.
Note that a perfect fluid with $p=k\rho$ ($0<k<1$) has a sound wave
at the speed $\sqrt{k}$, while a massless scalar field has 
a scalar wave at the speed of light.
This property results in a critical point of the ordinary differential
equations. This is called a sonic point for the perfect fluid case.
Critical points are classified through dynamical systems theory.
No information propagates inwardly beyond the sonic point.

\subsection{Global attractor and cosmic censorship}
\begin{figure}[htbp]
\begin{center}
\includegraphics[width=5cm]{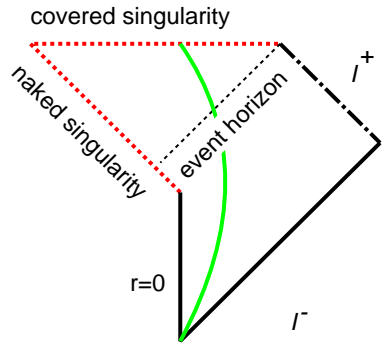}
\caption{\label{penrose_grlp} The Penrose diagram of the GRLP
solution for $0<k<0.0105$. The dashed line denotes the event horizon.}
\end{center}
\end{figure}

We focus on self-similar solutions with a perfect fluid with 
$p=k\rho$ ($0<k<1$).
Because of the critical point in the ordinary differential equations,
the smoothness condition strongly restricts the class of solutions.
We assume analytic initial data for self-similar perfect fluid solutions.
Analyticity in the present context means the Taylor-series expandability
of the energy density with respect to the Riemannian normal coordinates.
In particular, we impose analyticity both at the centre and at the sonic point.
Then we have only a discrete set of solutions.
They are the flat Friedmann solution, general relativistic Larson-Penston (GRLP) solution,
general relativistic Hunter (a) (GRHA) solution and so on. 
These are obtained numerically except for the flat Friedmann solution.
For these self-similar solutions, a set of analytic initial data is prepared 
at $t=t_{0}<0$ and singularity appears at $t=0$.
It is found that, except for the flat Friedmann solution,
the singularity appears at $t=0$ only at the centre and
we can analytically extend the solution beyond $t=0$ to positive $t$ for $r>0$.
If it is extended, the GRLP solution describes the formation of naked singularity from 
analytic and therefore regular initial data for $0<k<0.0105$~\cite{op1987,op1990}.
See Fig.~\ref{penrose_grlp} for the Penrose diagram of this
spacetime. We can see that the singularity is not completely 
hidden behind the event horizon. 

Since the cosmic censorship conjecture censors the generic occurrence 
of naked singularity, it is important whether the naked-singular solution
is stable or not. The self-similar solution is given representatively as
$H_{\rm ss}$. We assume that a linearly perturbed solution from this
is given in the following form:
\begin{equation} 
h(\tau,z)=H_{\rm ss}(z)+ \epsilon e^{\lambda \tau} F(z),
\end{equation}
where $\tau=-\ln (-t)$ and $|\epsilon|\ll 1$.
We can get equations for linear perturbation from Einstein's equation.
We impose regularity condition both at the centre and the sonic point.
Then, we can determine the value of $\lambda$ as an eigenvalue problem.
Then it is found that the GRLP has no unstable mode,
the GRHA has one unstable mode and 
other solutions except the flat Friedmann have more than one unstable 
modes~\cite{kha1995,kha1999,maison1996,hm2001,bcgn2002,snajdr2006}. 

Since the sonic point is a one-way membrane for sound waves,
there can appear a different kind of instability~\cite{harada2001}.
If we inject density gradient discontinuity at the sonic point, we can find
this discontinuity evolves locally at the sonic point.
This perturbation mode is called a kink mode.
The stability against this discontinuity is completely determined by the class 
to which the sonic point belongs as a critical point.
Then, it is found that, against this specific mode,
the flat Friedmann solution is unstable for $0<k\le 1/3$ and 
stable for $1/3<k\le 1$,
the GRLP is stable for $0<k<0.036$ and unstable for $0.036\le k<1/3$,
and the GRHA is stable for $0<k<0.89$ and unstable for $0.89\le k\le 1$.

As seen above, we have a self-similar solution which has 
no unstable mode. This is the GRLP solution for $0<k< 0.036$.
In fact, a numerical relativity experiment 
strongly suggests that this is a global attractor~\cite{hm2001}.
In the numerical simulation, the collapse ends in singularity 
formation at the centre for a certain subset 
of initial data sets, where the initial data sets
were prepared without fine-tuning. 
Then, the profile of the density profile tends to evolve in a 
self-similar manner and agree very well 
with the GRLP solution.
It was confirmed that this convergence to the self-similar 
attractor solution does not depend on 
the detailed choice of the initial density profile.
The above is the results of numerical simulations 
for $0<k\le 0.03$.
Although the first numerical simulation was done by
the simple Misner-Sharp scheme, 
this result has been recently confirmed by a
much more elaborated numerical scheme code
with high resolution shock capturing, adaptive mesh refinement
and innovative treatment of vacuum exterior~\cite{snajdr2006}.

This strongly suggests that the cosmic censorship 
is violated within spherical collapse. This is because 
the GRLP solution describes naked singularity formation for 
$0<k<0.0105$. Hence, this example is one of the strongest
counterexamples against the cosmic censorship.
On the other hand, the GRLP solution would be unstable 
against nonspherical perturbation for $0<k<1/9$, as the 
direct consequence of the linear perturbation 
analysis~\cite{gundlach2002}.
There remains much to study in nonspherical collapse.

\subsection{Convergence and critical phenomena}

Suppose we have a generic one-parameter
family of initial data sets parametrised by $p$.
Then, there generally exists a threshold value 
$p^{*}$ for black hole formation.
A near-critical collapse first approaches a self-similar
solution and deviates away eventually.
This self-similar solution that sits at the 
threshold is called a critical solution.
The scaling law for the formed black hole
for supercritical collapse is given as
\begin{equation}
M_{\rm BH} \propto |p-p^{*}|^{\gamma},
\end{equation}
for $p\approx p^{*}$,where $\gamma$ is called a critical exponent.
This is observed only as a result of fine-tuning
the parameter $p$ to be $p\approx p^{*}$.
The critical solution and critical exponent do not depend
on the prepared one-parameter family of initial data sets,
which is called universality.
The above phenomena were first observed in numerical simulation by
Choptuik~\cite{choptuik1993} for the spherical collapse of 
a massless scalar field 
and are called critical phenomena. It is found
that critical phenomena are seen in 
a variety of systems, such as
a perfect fluid with
$p=k\rho$~\cite{bcgn2002,snajdr2006,nc2000b,ec1994}.
See~\cite{gundlach2003} for an extensive review on this subject.

The critical behaviour turns out to be well understood by 
renormalisation group approach~\cite{kha1995}.
We consider the space of functions of $z$.
We can regard this as the space of initial data sets.
There the critical solution $H_{\rm ss}$ is characterised as a fixed 
point in this space with a single unstable mode. 
This means that the fixed point has a stable manifold of codimension
one.
Let $\lambda(>0)$ and $F_{\rm rel}$ be the eigenvalue and eigenfunction
of this unstable mode. 
The one-parameter family of initial data sets corresponds to a curve 
in the space of initial data sets and therefore generically has 
intersection with the stable manifold.
The intersection is actually 
the initial data for the exact critical collapse, i.e. $p=p^{*}$, 
which we denote as $H_{\rm c}$.
The initial data set for near critical collapse is then given by
\begin{equation}
h(0,z)=H_{\rm init}(z)=H_{\rm c}(z)+\epsilon F(z),\quad \epsilon=p-p^{*},
\end{equation}
where $\tau$ is chosen to be 0 for the initial time.
For large $\tau$, the deviation from the critical solution is dominated
by the unique unstable mode of the fixed point. The 
near-critical solution is then approximated as
\begin{equation}
h(\tau,z)\approx H_{\rm ss}(x)+\epsilon e^{\lambda \tau}F_{\rm rel}(z).
\end{equation}
Assuming that the deviation becomes of order unity at the black hole 
formation, we get the scaling law for the black hole mass as
\begin{equation} 
M_{\rm BH}= O(r) = O(e^{-\tau}) \propto |p-p^{*}|^{\gamma}, 
\end{equation}
where $\gamma=1/\lambda$. 
In fact, for the system of a perfect fluid with $p=k\rho$, 
the GRHA solution has a single unstable mode and 
acts as a critical solution. The observed critical exponent 
in the numerical experiment of gravitational collapse 
agrees very well with the above expected value obtained 
from the eigenvalue analysis. 
From this point of view, the GRLP solution, which has no unstable
mode, corresponds to an attractor of codimension zero, i.e. 
a global attractor.

The critical behaviour itself has important implication to 
cosmic censorship. In the limit of $p\to p^{*}$ from the 
supercritical regime, we will have a black hole of arbitrarily
small mass, which can be regarded as a naked singularity because
the curvature strength near the black hole horizon scales as $1/M^{2}$.
More directly, it is found that 
the Choptuik critical solution, which is discretely self-similar,
actually has a naked singularity at the centre~\cite{mg2003}. 
However, it should be noted that this naked singularity is realised as a
consequence of exact fine-tuning and hence nongeneric.
This is very different from the naked-singular GRLP solution, 
in which naked singularity generically appears
because the solution acts as a global attractor.

Lastly, we briefly review the Newtonian collapse of 
isothermal gas in the present context.
This system is much simpler than but still very similar 
to the general relativistic 
system of a perfect fluid.
There are a discrete set of spherically symmetric 
self-similar solutions with analytic initial data, 
including a homogeneous collapse, Larson-Penston solution,
Hunter (a) solution and so on~\cite{larson1969,penston1969,hunter1977,ws1985}.
The kink mode was also studied in this system~\cite{op1988}.
The numerical experiment of gravitational collapse and the normal mode analysis
show that there exist both convergence 
and critical phenomena~\cite{mh2001,hms2003}. The Larson-Penston solution 
acts as a global attractor solution, while the Hunter (a) solution 
acts as a critical solution.
This example shows that general relativity is 
not very essential to the appearance of critical behaviour.

\begin{figure}[htbp]
\begin{center}
\includegraphics[width=7cm]{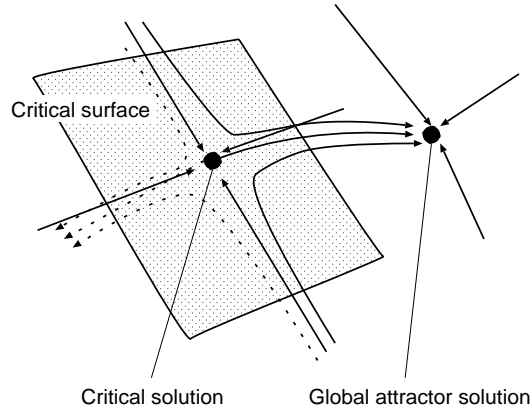}
\caption{\label{flow} The interplay of convergence and
 critical phenomena.}
\end{center}
\end{figure}

Figure~\ref{flow} schematically shows the renormalisation group flow
for this system. The global attractor solution (the Larson-Penston solution) 
has no unstable mode, while the critical solution (the Hunter (a) 
solution) has a single unstable mode.
A generic initial data set approaches the global attractor solution.
On the other hand, only if we tune the parameter near the critical
value, the supercritical initial data set first approaches the critical solution,
deviates away from it and then approaches the global attractor solution.
This interplay of the critical and global attractor solutions was first 
numerically observed in the Newtonian collapse of an 
isothermal gas~\cite{hms2003} 
and subsequently in the general relativistic system of a perfect fluid
$p=k\rho$ with sufficiently small positive $k$~\cite{snajdr2006}.

\section{Summary}
The spacetime singularity
has been a central issue in general relativity over several decades.
Recently numerical relativity has been developed and can reveal
the properties of generic spacetime singularities.
We have focused on the role of self-similar solutions
in singularity formation in gravitational collapse.
The general theory of relativity as well as Newtonian gravity
admits self-similar solutions. 
This is due to the scale-invariance of the theory.
The self-similar solutions are important not only because
they are dynamical and inhomogeneous solutions 
easier to obtain but also 
because they may play important roles 
in the asymptotic behaviour of more general solutions
in certain circumstances.
The numerical relativity experiments of gravitational collapse 
and the semi-analytical studies on self-similar solutions reveal 
that there is a self-similar solution which acts as a global attractor
in the spherical collapse of a perfect fluid and then the 
cosmic censorship will be violated if it is formulated within 
spherical symmetry.
We have also seen that the stability analysis of self-similar solutions 
gives a unified picture of the convergence to an attractor and the 
critical behaviour in gravitational collapse in terms of attractors
of codimension zero and one, respectively.
The interplay of these 
two behaviours has been recently
observed in numerical simulation of gravitational collapse 
both in Newtonian gravity and general relativity.
Since both the critical and convergence behaviours
are not only in general relativity but also in Newtonian gravity,
these two are considered to be common characteristics of gravitational 
physics and possibly of a wider class of scale-free nonlinear systems.

%%%%%%%%%%%%%%%%%%%%%%%%%%%%%%%%%
% References
%%%%%%%%%%%%%%%%%%%%%%%%%%%%%%%%%

\end{document}